\documentclass[twocolumn]{aastex631}

\usepackage{amsmath}

\definecolor{fuchsia}{rgb}{0.54, 0.17, 0.89}
\definecolor{azure}{rgb}{0.0, 0.5, 1.0}
\definecolor{pgreen}{rgb}{0.12, 0.3, 0.17}
\definecolor{alizarin}{rgb}{0.82, 0.1, 0.26}

\newcommand{\kms}{{\rm km~s^{-1}}}
\newcommand{\oii}{[\textrm{O}~\textsc{ii}]}
\newcommand{\oiii}{[\textrm{O}~\textsc{iii}]}

\newcommand{\oiiif}{$[\textrm{O}~\textsc{iii}]_{\lambda4363}$}
\newcommand{\nii}{[\textrm{N}~\textsc{ii}]}

\newcommand{\sii}{[\textrm{S}~\textsc{ii}]}
\newcommand{\ariii}{[\textrm{Ar}~\textsc{iii}]}
\newcommand{\ariv}{[\textrm{Ar}~\textsc{iv}]}

\newcommand{\simgt}{\,\rlap{\lower 3.5 pt \hbox{$\mathchar \sim$}} \raise
1pt \hbox {$>$}\,}
\newcommand{\simlt}{\,\rlap{\lower 3.5 pt \hbox{$\mathchar \sim$}} \raise
1pt \hbox {$<$}\,}

\newcommand{\ha}{${\rm H\alpha}$}
\newcommand{\hb}{${\rm H\beta}$}

\newcommand{\Z}{{\rm [Z/H]}}

\submitjournal{ApJ}

\shorttitle{Nitrogen abundances}
\shortauthors{Stiavelli, Morishita,  et al.}

\graphicspath{{./}{figures/}}

\begin{document}

\title{
What can we learn from the Nitrogen abundance of High-z galaxies?
}

\correspondingauthor{Massimo Stiavelli}
\email{mstiavel@stsci.edu}

\author[0000-0001-9935-6047]{Massimo Stiavelli}
\affiliation{Space Telescope Science Institute, 3700 San Martin Drive, Baltimore, MD 21218, USA}
\affiliation{The William H. Miller III, Dept. of Physics \& Astronomy, Johns Hopkins University, Baltimore, MD 21218, USA}
\affiliation{Dept. of Astronomy, University of Maryland, College Park, MD 20742, USA}

\author[0000-0002-8512-1404]{Takahiro Morishita}
\affiliation{IPAC, California Institute of Technology, MC 314-6, 1200 E. California Boulevard, Pasadena, CA 91125, USA}

\author[0000-0003-1564-3802]{Marco Chiaberge}
\affiliation{Space Telescope Science Institute for European Space Agency (ESA), ESA Office, 3700 San Martin Drive, Baltimore, MD 21218, USA}
\affiliation{The William H. Miller III, Dept. of Physics \& Astronomy, Johns Hopkins University, Baltimore, MD 21218, USA}

\author[0000-0003-4570-3159]{Nicha Leethochawalit}
\affiliation{National Astronomical Research Institute of Thailand (NARIT), Mae Rim, Chiang Mai, 50180, Thailand}

\author[0000-0002-5222-5717]{Colin Norman}
\affiliation{Space Telescope Science Institute, 3700 San Martin Drive, Baltimore, MD 21218, USA}
\affiliation{The William H. Miller III, Dept. of Physics \& Astronomy, Johns Hopkins University, Baltimore, MD 21218, USA}

\author[0000-0003-4223-7324]{Massimo Ricotti}
\affiliation{Dept. of Astronomy, University of Maryland, College Park, MD 20742, USA}

\author[0000-0002-4140-1367]{Guido Roberts-Borsani}
\affiliation{Department of Astronomy, University of Geneva, Chemin Pegasi 51, 1290 Versoix, Switzerland}

\author[0000-0002-8460-0390]{Tommaso Treu}
\affiliation{Department of Physics and Astronomy, University of California, Los Angeles, 430 Portola Plaza, Los Angeles, CA 90095, USA}

\author[0000-0002-5057-135X]{Eros~Vanzella}
\affiliation{INAF -- OAS, Osservatorio di Astrofisica e Scienza dello Spazio di Bologna, via Gobetti 93/3, I-40129 Bologna, Italy}

\author[0000-0002-4013-1799]{Rosemary F. G. Wyse}
\affiliation{The William H. Miller III, Dept. of Physics \& Astronomy, Johns Hopkins University, Baltimore, MD 21218, USA}

\author[0000-0003-3817-8739]{Yechi Zhang}
\affiliation{IPAC, California Institute of Technology, MC 314-6, 1200 E. California Boulevard, Pasadena, CA 91125, USA}

\author[0000-0003-4109-304X]{Kit Boyett}
\affiliation{Department of Physics, University of Oxford, Denys Wilkinson Building, Keble Road, Oxford OX1 3RH, UK}



\begin{abstract}

We present measurements of the gas-phase Oxygen and Nitrogen abundances obtained by applying the direct method to JWST NIRspec $R\sim1000$ spectroscopy for 6 galaxies at redshift greater than 3. Our measurements are based on rest-frame optical Nitrogen \nii$_{\lambda\lambda6548,6583}$ lines and are complemented by 6 additional objects from the literature at $3\leq z \leq 6$. We find that 9 out of 12 objects have values of log(N/O) that are compatible with 
those found for low-redshift, metal-poor, dwarf galaxies and for HII regions of more luminous local galaxies. However, 3 out of 12 objects have log(N/O) values that are overabundant compared to what is expected on the basis of their Oxygen abundance. We explore a few standard scenarios to explain the observations and conclude that, within the limited statistics available to us, none of them can be definitely excluded even though we prefer dilution by pristine gas infall in between star formation bursts as this is predicted by simulations to take place as a natural part of bursty star formation.

\end{abstract}

\keywords{}


\section{Introduction} \label{sec:intro}

For decades studies of high-redshift galaxies looked at chemical evolution as a one- or two- parameters story, depending on whether or not one could separate the enrichment due to Type II and Type Ia supernovae. However, the detailed history of chemical evolution in a galaxy is much more complex even though, until recently, it was hidden to our view. Today JWST \citep{gardner23} enables us to carry out detailed  galaxy interstellar medium studies  at $z\simgt4$ 
\citep{cameron23,Curti2023,Heintz2023,Laseter2023,nakajima23,sanders23,Shapley23,RobertsBorsani24,morishita24b}. The detailed line information allows us to determine the electron temperature and thus to measure abundances using the intrinsically more accurate ``direct" method as opposed to strong-line based abundance estimators. 

In addition to the gas-phase Oxygen abundance, we can now also probe abundance ratios \citep[e.g.,][]{morishita2024c}, and we are beginning to discover significant departures from the trends observed in objects in the local Universe, such as the reported over-abundance of Nitrogen in GNz11 \citep{bunkerGNz11,cameron23,senchyna23,Charbonnel2023} and in other galaxies \citep{MarquesChaves2024,Schaerer2024,topping24a,Topping2024b,Ji2024,isobe2023} or under-abundance of Carbon in MACS1149-JD1 \citep{stiavelli23}, the $z=6.23$ galaxy by \citet{jones23}, and MACS0647-JD \citep{Tiger24}. Probing different elemental abundances can reveal information on the detailed galaxy formation process, as stars of different masses, and hence different lifetimes, are the main producers of different elements. Thus, abundance ratio can become effective clocks to study early galaxy formation \citep{Tinsley1979,Matteucci1986,Gilmore1989}. Gas cycles, burstiness of star formation, and possibly exotic stellar populations or AGN can complicate the picture but we are now able to begin exploring these issues.

In this study, we combine our Cycle~1 \citep[PID:1199;][]{stiavelli23} and Cycle-2 (PID:2758) GTO programs to obtain gas-phase $N/O$ measurements using rest-frame optical lines for a small sample of galaxies at $z>3$. These datasets of sensitive spectra enable us to use lines for Nitrogen, Oxygen, and Hydrogen from similar ionization regions, ideally making the results more robust to ionization correction uncertainties. Our goal is to verify whether $N/O$ abundance ratios can provide clues on high-redshift galaxy formation as well as verifying whether optical rest-frame Nitrogen lines (statistically) provide similar results as rest-frame UV Nitrogen lines.

The paper is structured as follows:
In Sec.~\ref{sec:data} we present our data reduction, followed by our spectroscopic analysis in Sec.~\ref{sec:ana}. We characterize the resulting $N/O$ abundance ratios in Sec.~\ref{sec:res} and discuss the galaxy formation implications in Sec.~\ref{sec:disc}. Where relevant, we adopt the AB magnitude system \citep{oke83,fukugita96}, cosmological parameters of $\Omega_m=0.3$, $\Omega_\Lambda=0.7$, $H_0=70\,\kms\, {\rm Mpc}^{-1}$, and the \citet{chabrier03} initial mass function. For the solar abundances we follow \citet{Asplund2021}.

\section{Data and data reduction} \label{sec:data}

The data and their reduction for PID:1199 have already been reported in \citet{morishita24b}. PID:2758 consists of NIRSpec/MSA observations of galaxies in the NIRCam parallel field to PID:1199 i.e. ``1199-Parallel" (11:49:47.31, +22:29:32.1). These observations use the same gratings as PID:1199, and thus are configured with two medium-resolution gratings (G235M and G395M), aiming to capture rest-frame optical emission lines of galaxies at $z\sim3$--9. Unlike the case for PID:1199 (that had been planned pre-JWST commissioning before telescope pointing accuracy and stability were determined) these NIRSpec observations were executed over a single visit allowing for lower overhead. The position angle was set to PA=264.088\,degree. The total science time, excluding overhead, is 16806\,sec for each grating, {\it i.e.} the same within 1\% of the integration for PID:1199.

The MSA targets were selected on the basis of NIRCam photometry \citep{morishita24b} focusing on high-redshift star-forming galaxy candidates. The priority is given to $z>7$ galaxies (NIRCam F090W dropouts), $z>5$ (ACS F775W dropouts), and $z>2.4$ photometric-redshift sources. Potential AGN candidates at $z>3$ are also included in the MSA target list. The line fluxes we measure for the PID:2758 sample objects are given in Table~\ref{tab:linefluxes}.

For the MSA data reduction, we use {\tt msaexp}\footnote{https://github.com/gbrammer/msaexp} following \citet{morishita23b}. The one-dimensional spectrum is extracted via optimal extraction. The source light profile along the cross-dispersion direction is directly measured using the 2-d spectrum. For sources with faint continuum or with any significant contamination (i.e. from a failed open shutter), we visually inspect the 2-d spectrum and manually define the extraction box along the trace where any emission lines are identified.

\section{Spectroscopic analysis} \label{sec:ana}

We follow an approach similar to \citet{morishita24b} which we briefly describe here. We use sulphur-based densities whenever they are available (see Table~\ref{tab:phys}) and assume an electron density $n_e=300\,{\rm cm^{-3}}$ when they are not. Adopting $n_e=300\,{\rm cm^{-3}}$ for all objects would not yield significantly different results. We then derive an electron temperature using the direct method and the ratio of \oiiif\ to \hb\ \citep{izotov06}. The O$^{+}$ temperature is derived using \citet{Campbell86} and the oxygen metallicity is derived following \citet{izotov06}. We rely on \citet{izotov06} to derive the Nitrogen over Hydrogen ratio. We then derive the total Nitrogen over Oxygen ratio by applying the \citet{Amayo2021} Ionization Correction Factor (ICF) to ${\rm log}({\rm N}^+{\rm /O}^{+})$. This ICF is smaller than that given by \citet{izotov06} because it modifies the ratio of Nitrogen N$^+$ to O$^+$ which has a similar ionization energy. The \citet{izotov06} ICF to the ratio of Nitrogen-to-total Oxygen gives generally similar results but it requires much larger correction factors because total Oxygen is dominated by O$^{++}$ which has a very different ionization energy. 
From our GTO programs, we find 4 (1199) and 2 galaxies (2758) with spectra that enable Nitrogen abundance measurements.
We also derive Ar/O using [ArIV]$\lambda4740$ and/or [ArIII]$\lambda7135$  following \citet{izotov06} with Ionization Correction Factors from \citet{Amayo2021}. 

We have complemented our sample with the Oxygen and Nitrogen measurements by \citet{sanders23} for $5$ galaxies at $3 \leq z \leq 6$. We have used their published line fluxes and  re-analyzed them similarly to our data for homogeneity. The Oxygen abundance we derive for these objects are within the error bars of the values published by \citet{sanders23}. We have also added to the sample the galaxy Q2343-D40 by \citet{Rogers2024}. The quantities derived by our analysis as well as some physical parameters for the galaxies in the sample are listed in Table~\ref{tab:phys}. Star formation rates and stellar masses are from \citet{Morishita2024d}.

\begin{figure*}
\centering
	\includegraphics[trim={0 6cm 0 3cm},clip,width=0.95\textwidth]{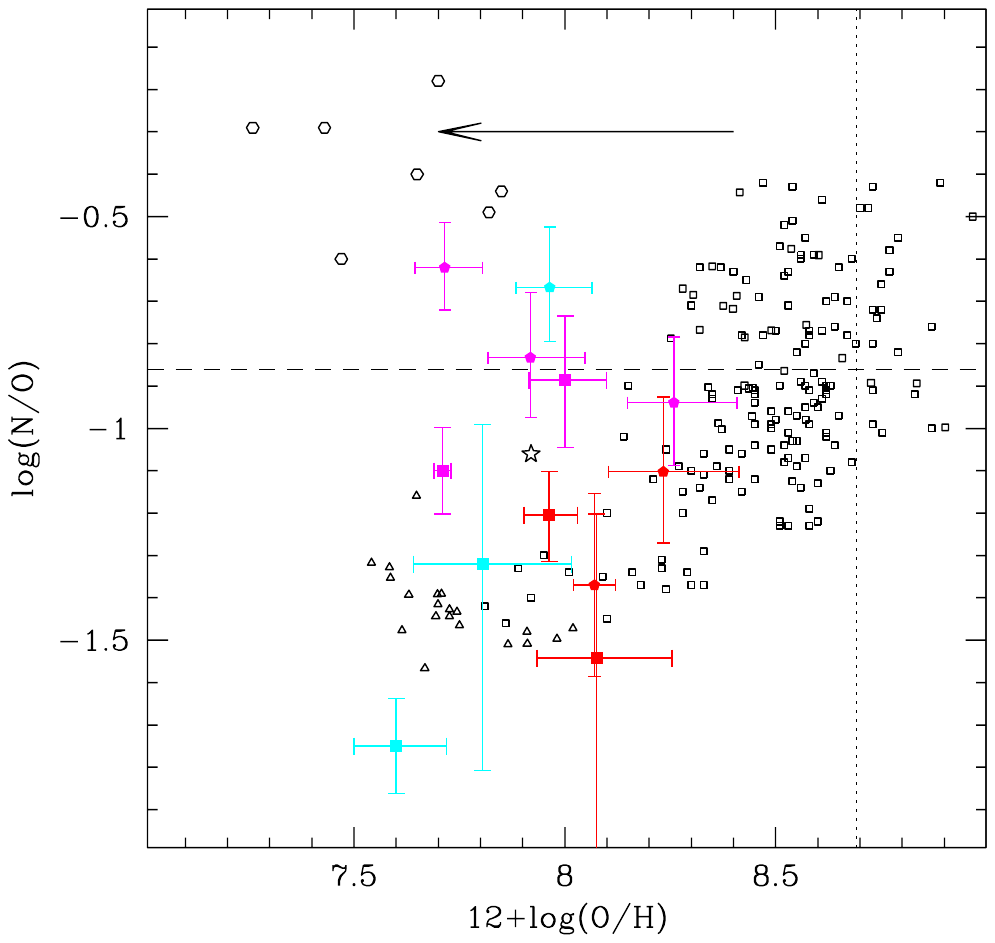}
	\caption{
 Measurements of log(N/O) vs 12+log(O/H)  for our sample galaxies (filled squares) and \citet{sanders23,Rogers2024} sample (small filled pentagons) compared to local dwarf galaxies from \citet{berg2019} (open triangles) and local extra-galactic HII regions from \citet{berg2020} (open squares). Rest-frame UV measurements from the literature are shown as open hexagons. The star indicates the Wolf-Rayet galaxy by \citet{morishita2024c}. The dashed and dotted lines represent the solar values for (N/O) and 12+log(O/H), respectively.
 Most high-redshift galaxies occupy the same region as local dwarfs but some show an excess of Nitrogen for their total Oxygen abundance. The redshift of the objects is color-coded: objects at $z<4$ are in red, those with $4\leq z<5$ in purple and those with $z\geq 5$ in cyan. The arrow at the top illustrated the effect of a dilution by a factor 5 \citep{sugimura24,pallottini2024}. 
    }
\label{fig:fig1}
\end{figure*}

\section{Results} \label{sec:res}

In Fig.~\ref{fig:fig1} we show the distribution of the sample galaxies in the log(N/O) vs 12+log(O/H) plane compared to local measurements \citep{berg2019,berg2020}. Three out of the twelve sample galaxies are significantly more enriched in Nitrogen than their gas-phase Oxygen abundance would suggest. An excess of Nitrogen has been observed in other galaxies on the basis of UV Nitrogen (but higher ionization) lines, N~III]$\lambda\lambda1747,1749$ and N~IV]$\lambda1486$  \citep{bunkerGNz11,isobe2023,MarquesChaves2024,Schaerer2024,topping24a,Topping2024b}. These objects are displayed as open hexagons. For the sake of homogeneity in the sample, we limit the following analysis to rest-frame optical Nitrogen \nii$_{\lambda\lambda6548,6583}$.

\begin{figure}
\centering
	\includegraphics[trim={0 5cm 0 5cm},clip,width=0.55\textwidth]{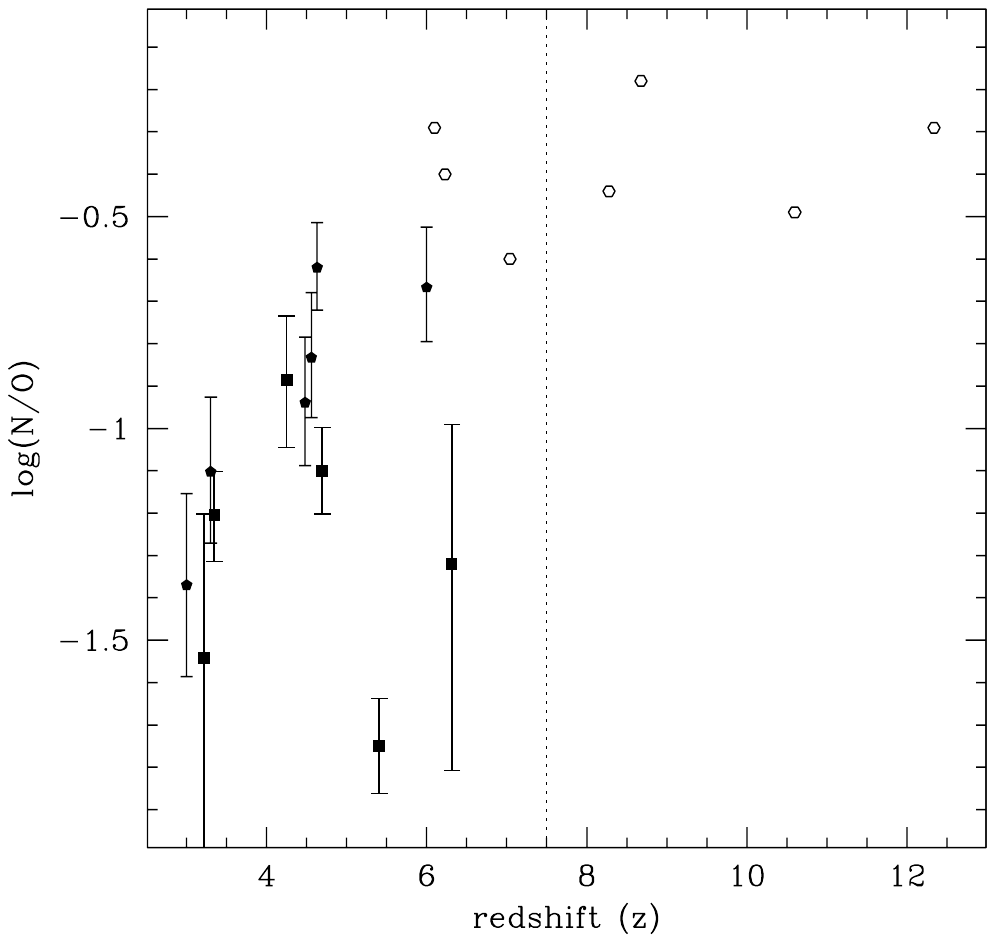}
	\caption{
log(N/O) vs redshift for galaxies in our sample. The rest-frame UV Nitrogen measurements from the literature are given in open hexagons. The dotted line shows the maximum redshift for measuring optical rest-frame Nitrogen lines with the NIRSpec instrument.   
    }
\label{fig:Nz}
\end{figure}

As a first test we have verified whether the Nitrogen overabundance might correlate with redshift. We find no strong evidence for that. 
The four galaxies at $z<4$ in our sample are consistent with the local samples, while one of the 3 at $z>5$ are overabundant in Nitrogen for their Oxygen abundance. We show in Fig.~\ref{fig:Nz} the values of log(N/O) vs redshift. It is interesting that the optical rest-frame Nitrogen measurements approach the UV rest-frame measurements at higher redshift. However, we see objects at the same redshift with lower values of log(N/O).

Possible, non-exotic, explanations for a Nitrogen overabundance compared to local dwarfs \citep[see, {\it e.g.}][]{morishita2024c}, are: {\it i)} infall of pristine gas diluting all abundances but not affecting abundance ratios, {\it ii)} effect of nitrogen enrichment by Wolf-Rayet stars, or {\it iii)} effect of Type II Supernova winds reducing Oxygen metallicity before intermediate mass stars enrich the gas with Nitrogen. 

In scenario {\it i)} it would not be natural to obtain $N/O$ values higher than those seen for the local sample as $N/O$ is unaffected and only the observed gas-phase Oxygen abundance is temporarily lowered by the infall. It is unlikely that this mechanism would be able to produce the $N/O$ values reported on the basis of rest-frame UV lines (the open hexagons in Fig.~\ref{fig:fig1}). On the other hand, this type of dilution is seen in simulations \citep{pallottini2024,sugimura24}, which could locate chemically enriched galaxies onto a low-O/H range and make them appear ``N/O-enriched" compared to their measured Oxygen abundance. 
The arrow in Fig.~\ref{fig:fig1} show the effect of dilution by primordial gas by a factor of 5 which is compatible with the simulations \citep{sugimura24,pallottini2024}.

In scenario {\it ii)}, the Nitrogen overabundance is a function of the timing of the observations with respect to the (bursty) star formation history. This has been modelled by \citet{kobayashi2024A} for the case of GNz11. In Fig.~\ref{fig:fig1} we show as a star the Wolf-Rayet galaxy of \citet{morishita2024c}. This galaxy displays only a modest Nitrogen overabundance and is not formally part of our sample because its redshift is below 3.

In scenario {\it iii)}, the Nitrogen overabundance is again due to bursty star formation, likely in clumps, where we can see objects that have ejected their Oxygen in SN-driven winds but retained their Nitrogen. 

\begin{figure}
\centering
	\includegraphics[trim={0 5cm 0 5cm},clip,width=0.55\textwidth]{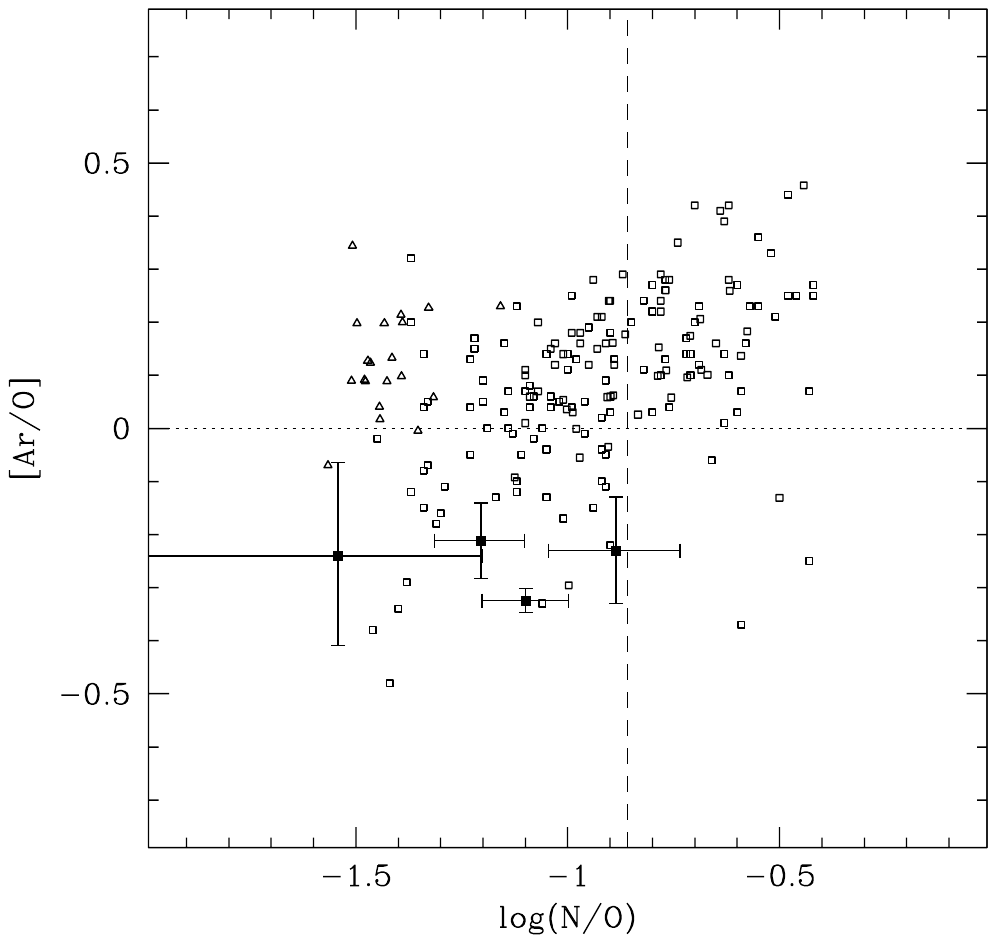}
	\caption{
[Ar/O] vs log(N/O) for galaxies in our sample compared to local dwarf galaxies from \citet{berg2019} (open triangles) and local HII regions from \citet{berg2020} (open squares). The dashed and dotted lines indicate the solar values for log(N/O) and [Ar/O] respectively.
    }
\label{fig:noAr}
\end{figure}

We have looked at the Argon abundance to see whether it could help shed light on the various scenarios.  Recent measurements of Argon abundances in high-$z$ galaxies have shown a distribution mostly compatible with local dwarf galaxies \citep[e.g.][]{Bhatt24} and these values support that there are multiple channels for Argon production including Type Ia supernovae. We have 4 measurements of Argon for galaxies with measured Nitrogen and they are all roughly similar, compatible with the lower envelope of the local dwarfs (see Fig.~\ref{fig:noAr}), and also compatible with objects that have not been enriched by Type Ia supernovae. Unfortunately, only one of the 4 objects has $N/O$ marginally over solar and that object is roughly compatible with the local sample anyway. Thus, also the Argon abundance analysis for our sample is not conclusive.

\section{Discussion and Conclusions} \label{sec:disc}

Using a homogeneous set of rest-frame optical line measurements we identify high-redshift galaxies whose ISM have higher Nitrogen abundance than expected on the basis of their gas-phase Oxygen abundance and the trends observed for local dwarf galaxies. All three galaxies that are overabundant compared to the local sample are at redshift $z>4.5$. On the other hand, four of the seven galaxies at $z>4.5$ are compatible with the local sample. Thus it is hard to claim conclusively that Nitrogen overabundance is a feature of high-z galaxies, even though we do not see it in any galaxy at $z<4.5$ in our sample 

\begin{figure}
\centering
	\includegraphics[trim={0 5cm 0 5cm},clip,width=0.55\textwidth]{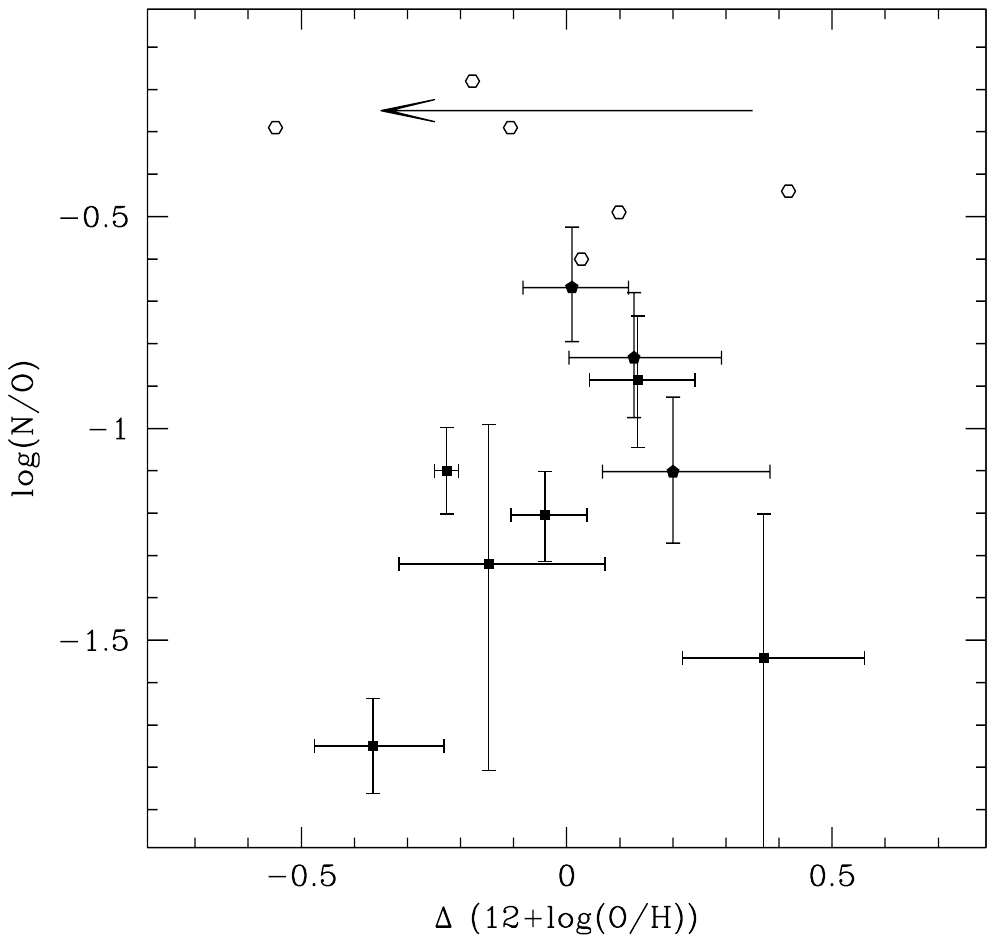}
	\caption{
We plot log(N/O) with respect to the Oxygen abundance $\Delta$ between the observed abundance and the one predicted on the basis of the galaxy stellar mass. Galaxies with gas-phase metallicity that has been diluted would be expected to have a low Oxygen abundance with respect to their stellar mass but this is not observed. Open hexagons are the galaxies with Nitrogen abundance derived from rest-frame UV lines. The arrow shows the expected effect of dilution by a factor of five.
    }
\label{fig:DMetal}
\end{figure}

We have considered some of the standard explanations for this behavior as we discussed in Section~\ref{sec:res}. Both the dilution by pristine gas, scenario {\it i)}, and the supernova wind scenario {\it iii)} are compatible with our observations as the Nitrogen overabundant galaxies have low Oxygen abundance for their derived stellar mass.  The dilution scenario would require dilutions by factors 3-10 for the three galaxies with the highest values of ${\rm log}(N/O)$ ($\geq-0.85$). These factors are well within what can be expected on the basis of existing galaxy formation models \citet{sugimura24,pallottini2024}. One could expect galaxies with Nitrogen excess due to dilution (or winds) to show a metal deficit for their stellar mass, because these objects would have effectively lost or diluted their metal content. In \citet{morishita24b} we derive a mass-metallicity relation for galaxies in the redshift range 3 to 9.5 using direct abundance measurement using auroral lines as done here. What we refer to as "metallicity" in that paper is the gas-phase Oxygen abundance. By adopting tthis relation from \citet{morishita24b} we show in Fig.~\ref{fig:DMetal} how $N/O$ compares to the $\Delta (12+{\rm log}({\rm O/H})$ with respect to the correlation. In Fig.~\ref{fig:DMetal} we show that such an expected correlation is not observed as the N-overabundant galaxies in our sample are not under-abundant in Oxygen for their stellar mass.

In general we see high-ionization, as inferred from the \oiii/\oii\ line ratio, only when O stars are present. This makes
the supernova wind scenario slightly more contrived. In that scenario one would need a first burst creating the Type II SNae responsible for the wind clearing up the object from its O-enriched gas, followed by intermediate mass stars evolution producing Nitrogen. A final burst of star formation would produce the O-stars needed for ionizing the N-rich gas that we observe. Even in this scenario one would expect the Nitrogen overabundant galaxies to have low Oxygen abundance for their mass which is not observed (see Fig.~\ref{fig:DMetal}). The need for this specific timing of events makes this scenario less palatable than the dilution scenario.

We find the possibility that Nitrogen is temporarily enriched by Wolf-Rayet stars, scenario {\it ii)}, to be also less compelling because of the timing fine tuning required \citep{kobayashi2024A}.
However, given the small sample size we do not feel that we can rule it out. In fact, we think that none of the standard scenarios can be ruled out at this time.

More exotic explanations for the Nitrogen abundance at high redshift, as well as to explain the (N/O) ratios seen in Milky Way Globular Clusters, have been proposed \citep[e.g.][]{senchyna23,Charbonnel2023} but we do not think that they are necessary to explain our observations given that the Nitrogen overabundance is significant relative to the observed (low) gas-phase Oxygen abundance but is only comparable or modestly in excess of the solar [N/O] ratio. On the other hand, these explanations might well be required to explain the more significant Nitrogen overabundance derived from rest-frame UV lines. Indeed, it is hard to imagine that the standard scenarios we consider could explain the highest Nitrogen overabundance values derived from rest-frame UV lines. Could it be a single process producing the observed Nitrogen overabundance? Certainly, looking at Fig.~\ref{fig:Nz} would hint that that might be the case. On the other hand, one should be careful when comparing measurements based on lines with different excitation and possibly much higher electron densities \citep[e.g.][ see also Zhang et al. in preparation]{topping24a,Topping2024b,MarquesChaves2024}. 

Given these conclusions, further progress in this area will likely require: better statistics, increases in the number of objects for which both rest-frame UV and optical Nitrogen lines are available, and additional abundances that could be used as chemical clocks to distinguish the various scenarios. One such element could be Carbon.  We are looking at obtaining Carbon and rest-frame UV measurements for some of these objects in an upcoming Cycle 3 program (PID: 4552).

\begin{acknowledgments}

The data presented in this paper were obtained from the Mikulski Archive for Space Telescopes (MAST) at the Space Telescope Science Institute.  We acknowledge support for this work under NASA grant 80NSSC22K1294.

\end{acknowledgments}

This work is based in part on observations made with the NASA/ESA/CSA James Webb Space Telescope. The data were obtained from the Mikulski Archive for Space Telescopes at the Space Telescope Science Institute, which is operated by the Association of Universities for Research in Astronomy, Inc., under NASA contract NAS 5-03127 for JWST.  Support for program JWST-GTO2758 was provided by NASA through grant 80NSSC21K1294.
MS thanks A. Ferrara, A. Pallottini, A. Marconi for useful discussions. KB acknowledge funding from the "FirstGalaxies" Advanced Grant from the European Research Council (ERC) under the European Union’s Horizon 2020 research and innovation programme (Grant agreement No. 789056).
The specific observations analyzed can be accessed via \dataset[DOI: 10.17909/1d3v-5f80 ]{PID: 2758}. These observations are associated with program JWST-GTO1199 and GTO2758.

%

\vspace{5mm}
\facilities{JWST(NIRSpec)}


\software{
Astropy \citep{astropy13,astropy18,astropy22}, bbpn \citep{bbpn}, EAzY \citep{brammer08}, EMCEE \citep{foreman13}, gsf \citep{morishita19}, numpy \citep{numpy}, python-fsps \citep{foreman14}, JWST pipeline \citep{jwst}.
          }



\appendix

%


\begin{deluxetable*}{ccccccccccc}
\tabletypesize{\footnotesize}
\tabcolsep=0.09cm
\tablecolumns{11}
\tablecaption{
}
\tablehead{
\colhead{ID} & \colhead{\oii$_{3727}$} & \colhead{\oiii$_{4363}$} & \colhead{\ariv$_{4740}$} & \colhead{\hb$_{4861}$} & \colhead{\oiii$_{4959+5007}$} & \colhead{\nii$_{6548+6585}$} & \colhead{\ha$_{6563}$} & \colhead{\sii$_{6716}$} & \colhead{\sii$_{6731}$} & \colhead{\ariii$_{7136}$}
}
\startdata
\cutinhead{GTO~2758 Sample}
2758\_40171 & -- & {$<7.6$} & -- & {$52.9 \pm 3.5$} & {$367.7 \pm 14.3$} & {$<5.9$} & {$112.7 \pm 3.6$} & {$10.1 \pm 2.5$} & {$<4.8$} & {$<2.6$}\\
2758\_60001 & {$60.5 \pm 2.8$} & {$17.5 \pm 1.0$} & {$1.8 \pm 0.6$} & {$124.6 \pm 2.6$} & {$1033.8 \pm 29$} & {$11.8 \pm 0.5$} & {$475.4 \pm 7.6$} & {$5.7 \pm 0.6$} & {$3.9 \pm 0.5$} & {$5.2 \pm 0.7$}\\
2758\_20006 & {$<2.5$} & {$<2.8$} & {$<4.5$} & {$8.8 \pm 1.1$} & {$73.8 \pm 4.8$} & {$<2.3$} & {$26.9 \pm 1.1$} & {$<2.0$} & {$<1.9$} & {$<2.4$}\\
2758\_40149 & {$26.3 \pm 1.5$} & {$7.0 \pm 1.4$} & {$<3.9$} & {$36.4 \pm 1.4$} & {$323.6 \pm 6.8$} & {$30.4 \pm 4.2$} & {$172.3 \pm 2.8$} & {$<1.8$} & {$<2.2$} & {$<2.8$}\\
2758\_40004 & {$13.7 \pm 1.4$} & {$2.8 \pm 1.0$} & {$<1.5$} & {$9.6 \pm 1.3$} & {$90.1 \pm 7.6$} & {$<2.6$} & {$34.8 \pm 1.5$} & {$2.8 \pm 0.8$} & {$<2.6$} & {$<3.3$}\\
\cutinhead{GTO~1199 Sample}
1199\_30055 & -- & {$18.8 \pm 5.6$} & --  & {$118.4 \pm 5.4$} & {$1358.7 \pm 34.0$} & {$<16.1$} & {$438.3 \pm 6.7$} & {$10.2 \pm 3.3$} & {$6.9 \pm 2.0$} & --\\
1199\_150880 & {$88.1 \pm 4.1$} & {$11.9 \pm 2.1$} & -- & {$87.6 \pm 2.7$} & {$898.7 \pm 15.8$} & {$8.6 \pm 1.4$} & {$286.9 \pm 3.0$} & {$7.4 \pm 1.0$} & {$6.0 \pm 1.0$} & --\\
\enddata
\tablecomments{
Fluxes are in units of $10^{-19}$\,erg/s/cm$^2$. Flux errors are 1-$\sigma$. 2.5-$\sigma$ upper limits are quoted for non-detected lines (SN\,$<2.5$). Lines without the spectroscopic data coverage are marked with ``--". Among the 1199 samples, we present two that have [S~II] detections. The remaining objects are reported in \citet{morishita24b}.}\label{tab:linefluxes}
\end{deluxetable*}

\begin{deluxetable*}{cccccccccccccc}
    \tabletypesize{\footnotesize}
    \tabcolsep=0.09cm
    \tablecolumns{14}
    \tablewidth{0pt}
    \tablecaption{
    Physical properties of the samples
    }
    \tablehead{
    \colhead{ID} & \colhead{R.A.} & \colhead{Decl.} & \colhead{$m_{\rm F444W}$} & \colhead{$z$} &  \colhead{$n_e$} & \colhead{$\log$(O/H)} & \colhead{$T_e$} & \colhead{AV$_{\rm neb.}$} & 
    \colhead{$\log$(N/O)} & \colhead{$\log$(Ar/O)} &
    \colhead{$\log M_*$} & \colhead{$\log$\,SFR} & \colhead{$\mu$} \\
    \colhead{} & \colhead{degree} & \colhead{degree} & \colhead{mag} & \colhead{} & \colhead{cm$^{-3}$} & \colhead{+12} &  \colhead{$\times10^4$\,K} & \colhead{mag} & \colhead{} & \colhead{} & \colhead{$M_\odot$} & \colhead{$M_\odot/{\rm yr}$} & \colhead{}
    }
    \startdata
    \cutinhead{GTO~1199 Sample}
    30055& $177.388125$ & $22.415668$ & 24.9 & 3.214 & $1^{+1260}_{-1}$ & $8.07_{-0.14}^{+0.18}$ & $1.58$ & $0.98$ & $-1.53^{+0.34}_{-0.66}$ & $-2.55^{+0.18}_{-0.17}$ & $8.68_{-0.04}^{+0.04}$ & $0.6_{-0.1}^{+0.1}$ & $5.8_{-1.8}^{+5.9}$ \\
    20028& $177.410476$ & $22.419310$ & 23.6 & 3.345 & - & $7.96_{-0.06}^{+0.07}$ & $1.51$ & $0.84$ & $-1.20^{+0.10}_{-0.11}$& $-2.52^{+0.07}_{-0.07}$ & $9.79_{-0.04}^{+0.01}$ & $1.8_{-0.1}^{+0.1}$ & $1.4_{-0.1}^{+0.1}$ \\
    150880& $177.403656$ & $22.368355$ & 26.6 & 4.247 & $203^{+500}_{-203}$ & $8.00_{-0.08}^{+0.10}$ & $1.49$ & $0.50$ & $-0.88^{+0.15}_{-0.16}$& $-2.54^{+0.10}_{-0.10}$& $9.28_{-0.04}^{+0.02}$ & $0.9_{-0.1}^{+0.1}$ & $1.3_{-0.1}^{+0.1}$ \\
    320108& $177.402573$ & $22.384069$ & 23.7 & 4.257 & - & $8.36_{-0.05}^{+0.05}$ & $1.09$ & $1.16$ & - & - & $10.13_{-0.01}^{+0.01}$ & $1.6_{-0.1}^{+0.1}$ & $2.2_{-0.1}^{+0.1}$ \\
    320002& $177.428253$ & $22.409452$ & 27.5 & 4.658 & - & $7.41_{-0.16}^{+0.19}$ & $2.59$ & $1.24$ & - & - & $8.68_{-0.05}^{+0.08}$ & $0.7_{-0.1}^{+0.1}$ & $1.3_{-0.1}^{+0.1}$ \\
    150903& $177.424210$ & $22.408874$ & 26.4 & 4.659 & - & $7.90_{-0.09}^{+0.11}$ & $1.68$ & $0.00$ & - & - & $9.25_{-0.01}^{+0.01}$ & $0.9_{-0.1}^{+0.1}$ & $1.4_{-0.1}^{+0.1}$ \\
    10010& $177.412949$ & $22.418858$ & 25.7 & 6.311 & - & $7.81_{-0.16}^{+0.21}$ & $1.78$ & $0.62$ & $-1.32^{+0.33}_{-0.49}$& - & $9.59_{-0.03}^{+0.02}$ & $1.0_{-0.1}^{+0.1}$ & $1.4_{-0.1}^{+0.1}$ \\
    2& $177.417709$ & $22.417431$ & 23.8 & 7.232 & - & $7.92_{-0.04}^{+0.04}$ & $1.64$ & $0.00$ & - & - & $9.46_{-0.03}^{+0.02}$ & $1.9_{-0.1}^{+0.1}$ & $1.4_{-0.1}^{+0.1}$ \\
    3& $177.389908$ & $22.412710$ & 25.6 & 9.114 & - & $7.82_{-0.07}^{+0.07}$ & $1.66$ & $0.00$ & - & - & $7.77_{-0.02}^{+0.04}$ & $0.5_{-0.1}^{+0.1}$ & $12.5_{-5.0}^{+12.0}$ \\
    \cutinhead{GTO~2758 Sample}
    20006 & 177.433685 & 22.467369 & 27.9 & 5.229 & - & $7.32_{-0.19}^{+0.30}$ & $1.44$ & $0.26$ & - & - & $8.5$ & $0.2$ & - \\
    40004& 177.466064&  22.503035& 27.4& 5.682 & - & $7.50_{-0.18}^{+0.24}$ & $1.84$ & $0.90$ & - & - & $8.7$ & $0.6$ & - \\
    40149& 177.425964&  22.468485 & 25.4& 5.403 & - & $7.56_{-0.10}^{+0.12}$ & $2.09$ & $1.38$ & $-1.75^{+0.11}_{-0.11}$ & - & $9.6$ & $1.1$ & - \\
    40171& 177.435959& 22.469589& 25.9& 3.268 & - & $7.62_{-0.19}^{+0.28}$ & $1.20$ & $0.0$ & - & - & $9.2$ & $0.5$ & -\\
    60001& 177.465897&  22.518959 & 25.4& 4.694 & $261^{+210}_{-150}$ & $7.71_{-0.02}^{+0.02}$ & $1.76$ & $1.09$ & $-1.10^{+0.10}_{-0.10}$ & $-2.63_{-0.02}^{+0.02}$ & $9.5$ & $1.0$ & - \\
    \cutinhead{Sanders Sample}
    397& 214.83620& 52.882690& 25.2& 6.000 & - & $7.96_{-0.08}^{+0.10}$ & $1.43$ & $0.09$ & $-0.67^{+0.14}_{-0.13}$ & - & 9.63 & 1.63 & - \\
    1477& 215.003490& 52969540& - & 4.631 & - & $7.72_{-0.07}^{+0.09}$ & $1.84$ & $0.62$ & $-0.62^{+0.11}_{-0.10}$ & - & -$^\ast$ & -$^\ast$ & - \\
    1746& 215.05401& 52.956870& - & 4.569 & $790^{+2300}_{-590}$ & $7.92_{-0.10}^{+0.13}$ & $1.59$ & $0.0$ & $-0.85^{+0.15}_{-0.14}$ & - & 9.03 & 0.77 & - \\
    1665& 215.178200& 53.059350& -& 4.482 & $204^{+390}_{-204}$ & $8.26_{-0.11}^{+0.15}$ & $1.2$ & $0.89$ & $-0.93^{+0.15}_{-0.15}$ & - & -$^\ast$ & -$^\ast$ & - \\
    11088& 214.93421& 52.826370& - & 3.302 & $164^{+204}_{-138}$ & $8.23_{-0.13}^{+0.18}$ & $1.20$ & $1.21$ & $-1.10^{+0.17}_{-0.17}$ & - & 9.95 & 1.11 & -\\
    \cutinhead{Rogers Sample}
    Q2343-D40 &  &  &  & 2.963 & $73^{+71}_{-71}$ & $8.07_{-0.06}^{+0.06}$ & $1.32$ & $0.10$ & $-1.37_{0.21}^{+0.21}$ & $-2.80_{-0.12}^{+0.12}$ & - & - & - \\
    \enddata
    \tablecomments{
    $\dagger$: Only galaxies with direct abundance measurements via \oiiif\ are shown  Errors and upper limits are $1\,\sigma$.
    $\ast$: SED fit is not attempted due to the lack of JWST NIRCam data.
    Electron temperature ($T_e$) in units of $10^4$\,K. $\mu$ is the lensing amplification from \citet{morishita24b}. Stellar masses and SFR are from \citet{Morishita2024d}
    }\label{tab:phys}
    \end{deluxetable*}


\bibliography{out}{}
\bibliographystyle{aasjournal}



\end{document}